\documentclass{article}
\usepackage[paperwidth=8.5in,margin=.9in]{geometry}
\usepackage{amsmath,abstract,bm,subfig}
\usepackage{graphicx}
\usepackage{cite}
\usepackage[utf8]{inputenc}
\usepackage[english]{babel}
\usepackage{csquotes}

\newcommand\unx[1]{\text{#1}}
\newcommand\un[1]{\,\text{#1}}

\newcommand{\lr}[1]{\left( #1 \right)} 
\newcommand{\vc}[1]{{\bf #1}}
\newcommand{\vep}{\bm{\varepsilon}}
\newcommand{\E}{\bm{\mathcal{E}}}
\newcommand{\env}{\text{env}}
\DeclareMathOperator{\sinc}{sinc}
\DeclareMathOperator{\He}{He}

\begin{document}

\title{THz based phase space manipulation in a zero-slippage IFEL\thanks{Work supported by DOE grant DE-FG02-92ER40693 and NSF grant PHY-1415583.}}
\author{E. Curry\thanks{ejcurry@physics.ucla.edu}, S. Fabbri, P. Musumeci, UCLA, Los Angeles, CA 90049, USA\\
A. Gover, Tel-Aviv University, Tel-Aviv 69978, Israel}
\maketitle
\date{}

\begin{abstract}
We describe an IFEL interaction driven by a guided broadband THz source to compress a relativistic electron bunch and synchronize it with an external laser pulse. A high field near single-cycle THz pulse, generated via optical rectification from the external laser source, is group velocity-matched to the electron bunch inside a waveguide, allowing for a sustained interaction in a magnetic undulator. We present measurements of the THz waveform before and after a curved parallel plate waveguide with varying aperture size and estimate the reduced group velocity. For a proof-of-concept experiment at the UCLA PEGASUS laboratory, a $6\un{MeV}$, $100\un{fs}$ electron bunch with an initial $3\times10^{-4}$ energy spread can be readily produced. Given these parameters and a projected THz peak field of $10\un{MV/m}$, our simulation model predicts a phase space rotation of the bunch distribution that compresses the electron bunch by nearly an order of magnitude and reduces any initial timing jitter within the phase acceptance window. We also discuss the application of this guided-THz IFEL experimental set-up towards a THz streaking diagnostic with the potential for femtosecond scale temporal resolution.
\end{abstract}

\section*{Introduction}

The unique properties of the radiation in the THz frequency regime have made it an invaluable tool for imaging and spectroscopy \cite{spectroscopyJepsen}. Until recently, the progress of research on further THz applications in the accelerator community has been limited by the available power, but with improved generation techniques, such as pulse-front tilted optical rectification \cite{pulsefronttiltStepanov}\cite{pulsefronttiltHebling} and photoconductive terahertz optoelectronics \cite{pcaJarrahi}, THz sources have become an attractive tool for accelerator science. In particular, for beam phase space manipulation, the THz frequency range offers a unique middle ground between the high accelerating gradients of laser frequencies and the broad phase-acceptance window of radio frequency (RF) waves. For example, the strong energy chirp imparted to an electron bunch in an X-band structure, like the one used for phase space linearization at LCLS \cite{LCLScompressionEmma}, could be accomplished by a THz field that is over fifty times smaller because of the higher frequency. At the same time, where laser coupling in a typical free electron laser (FEL) results in a train of microbunches, a THz-driven FEL interaction would capture and compress the entire beam in a single ponderomotive bucket.

To couple the THz wave to the longitudinal beam momentum, one can induce a longitudinal component in the field or a transverse component in the electron trajectory. Given excitation of the appropriate mode, a dielectric or corrugated waveguide structure can provide a longitudinal field for direct coupling to the beam, as demonstrated in ref.~\cite{THzlinacNanni} for low energy electrons. In an inverse free electron laser (IFEL), a series of alternating polarity dipole magnet pairs (an undulator) is used to produce transverse oscillations in the beam trajectory to enable ponderomotive coupling with a transverse THz field. While the peak power available from today's THz sources continues to improve, it remains the limiting factor for most THz accelerator applications. The IFEL coupling scheme has unique advantages which allow it to harness the available power more effectively, as we will see below, and is the basis for our THz-based beam manipulation technique.

At the UCLA PEGASUS laboratory, we intend to demonstrate the compression of a low charge electron bunch in a magnetic undulator using ponderomotive coupling with a THz pulse. A single Ti:Sapph laser source will be used to generate an intense, nearly single-cycle THz wave through optical rectification while simultaneously driving a 1.6 cell S-band photocathode gun. The ponderomotive force produced by the THz and undulator fields induces a strong energy chirp on the electron bunch, which can be converted into longitudinal compression after a drift section. To sustain the IFEL-type interaction over the length of the undulator, the group velocity of the THz pulse will be matched to the propagation of the beam using a curved parallel plate waveguide to produce a "zero-slippage" resonance condition.

In addition to strong compression, the interaction reduces the time-of-arrival jitter of the electron bunches relative to the laser pulse used to generate them. When the phase of the ponderomotive bucket is centered on the average arrival time of the electron bunches, the induced energy chirp works to accelerate late bunches and decelerate early bunches towards the optimal timing as the bunch is compressed. This effect is relevant for all applications involving laser-electron interactions which have tight timing tolerances such as pump-probe techniques~\cite{pumpprobeGlownia}, Inverse Compton Scattering sources~\cite{ICSscalingHartemann} and external injection into advanced accelerators~\cite{LWFAreviewEsarey}.

This paper is organized as follows. We begin with a discussion of the advantages of the IFEL ponderomotive coupling and of the zero-slippage interaction for beam compression. We then present experimental results demonstrating the dispersive properties of the curved parallel plate waveguide, suitable for group-velocity matching of the THz pulse with the electron beam. An electro-optic sampling technique is used to characterize the dispersed radiation pulse profile. We include simulation results of the guided-THz IFEL interaction using the beam parameters at PEGASUS laboratory, and conclude with a discussion of an opportunity offered by the guided-THz IFEL setup to perform high-streaking-speed THz-based temporal diagnostics.

\section*{Zero-slippage IFEL interaction}

The main advantage of transverse ponderomotive coupling using an undulator magnetic field (commonly referred to as FEL or IFEL coupling) with respect to longitudinal "slow-wave" (Cherenkov Smith-Purcell) coupling in a dielectric or corrugated waveguide structure is related to the transverse acceptance of the interaction region. This has been discussed in detail in the extensive comparison of laser to beam coupling schemes compiled in ref.~\cite{FELtechniquesGover} which we briefly review here to highlight the result.

In order to enable efficient energy transfer in an IFEL interaction between the electrons and a wave of frequency $\omega$ and wavenumber $k_z$, the phase synchronism condition 
\begin{equation}
\frac{\omega}{c \beta_z} = k_z+k_u
\label{phase_synchronism}
\end{equation}
must be satisfied, where $\beta_z$ is the normalized longitudinal component of the electron velocity in the undulator, $k_u$ is the wavenumber describing the periodicity of the undulator, and for free-space propagation $k_z = \omega /c$. For longitudinal coupling, the dielectric or the structure corrugation period are chosen so that there exists one waveguide mode that satisfies the phase synchronism condition $\frac{\omega}{c \beta_z} = k_z$.

For both IFEL and longitudinal coupling schemes, the periodic magnetostatic field or slow radiation mode decays, in terms of the distance, $x$, away from the magnetic field source or slow wave structure wall, like $e^{-k_{\perp} x}$. The decay constant, $k_{\perp}$, is determined by the geometry of the periodic field source or slow wave structure and by the interaction frequency $c/\lambda$. Comparing the cases of transverse coupling in an undulator and longitudinal coupling with the TM mode in a guiding structure we have 

\begin{align}
k_{\perp} = \Bigg{\{}
\begin{matrix}
\renewcommand*{\arraystretch}{2}
 k_u \text{\, ,} & \text{transverse coupling} \\ \lr{k_z^2-\frac{\omega^2}{c^2}}^{1/2}\text{\, ,} & \text{longitudinal coupling.}
\end{matrix}
\end{align}

For the case of transverse ponderomotive coupling in an undulator, $k_{\perp}$ must be equivalent to the undulator wavenumber $k_u$ to satisfy the magnetostatic Laplace equation. In the case of longitudinal coupling with a TM mode, $k_{\perp}$ should satisfy the Helmholtz equation. We can define an interaction range, $\frac{1}{2k_{\perp}}$, in which there is strong coupling for beam-wave interaction. Under the condition of phase synchronism \eqref{phase_synchronism}, this interaction range can be expressed in terms of the beam Lorentz factor and normalization by the wavelength for both schemes:
\begin{align}
\frac{1}{2k_{\perp}} = \frac{\lambda}{4\pi} \Bigg{\{}
\begin{matrix}
\renewcommand*{\arraystretch}{2}
 \frac{\beta_z}{1-\beta_z}\text{\, ,} & \text{transverse coupling} \\ \beta_z \gamma_z\text{\, ,} & \text{longitudinal coupling.}
\end{matrix}
\end{align}

In the non-relativistic limit, the interaction ranges have equivalent magnitudes for both manipulation techniques, but as one approaches the highly relativistic limit the interaction range is proportional to $2\gamma_z^2$ for the transverse coupling technique and to $\gamma_z$ for the longitudinal technique. This is an important point when the wavelength of the accelerator is shrunk by two or three orders of magnitude from conventional radiofrequency to the THz range. From this analysis, we find that, for relativistic beams, the cross-section of the beam accelerated by the THz wave can be significantly larger in the undulator coupling case. 

These considerations assumed dispersion-free vacuum propagation of the radiation for the undulator coupling, but yield equivalent results if one chooses to use a waveguide TE mode for the radiation in the undulator coupling. Note also that the TE modes used for transverse coupling typically suffer from less attenuation within the guiding structure than the TM modes in a dielectric or corrugated waveguide structure~\cite{FELtechniquesGover,waveguidestudyYakover}. For these reasons the use of a wiggler structure interaction scheme (IFEL) is advantageous over a slow wave structure interaction for the purpose of bunching relativistic beams.

The IFEL resonance condition for free-space interaction assumes that the radiation slips ahead of the particles by an integer number of wavelengths for every undulator period to maintain phase synchronism along the undulator. The near single-cycle THz pulse produced by optical rectification precludes the use of this slippage mechanism to keep phase synchronism. Therefore, in order to sustain an interaction between a near single-cycle THz pulse and electron beam in the undulator, one must operate in the "zero-slippage" condition, in which a waveguide is used to match the THz group velocity to the average longitudinal speed of the electron bunch \cite{zeroslippageDoria}\cite{zeroslippageGover}.

\begin{figure}[!htb]
   \centering
   \includegraphics*[scale=.5]{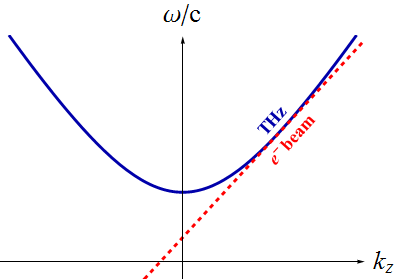}
   \caption{The dispersion curves for the waveguide (blue) and electron beam (dashed red). The "zero-slippage" condition occurs when their slopes, corresponding to radiation group velocity and average longitudinal bunch velocity, are equal.}
   \label{zeroslippageplot}
\end{figure}

Group and phase velocity synchronism occur simultaneously when the dispersion curve of a waveguide mode, $k=k_{zq}(\omega)$ with mode index $q$, and the displaced dispersion relation of the e-beam modulation wave, $k_z=\frac{\omega}{c \beta_z} -k_u$, intersect tangentially at the THz frequency, $\omega_0$, as shown in Fig.~\ref{zeroslippageplot}. The group velocity synchronism condition, stated as
\begin{align}
v_g=\frac{d\omega}{dk_z}\Bigg|_{\omega_0} = c \beta_{z}
\label{vel_res}\text{,}
\end{align}
is satisfied for frequencies at which the curves in Fig.~\ref{zeroslippageplot} have the same slope, while the phase synchronism condition is satisfied for frequencies at which the curves intersect. For the case of tangential intersection, group and phase velocity synchronism is achieved at $\omega=\omega_0$, and nearly so in a wide range of frequencies around $\omega_0$.

Simultaneous solution of both conditions, \eqref{phase_synchronism} and \eqref{vel_res}, can be found explicitly when the dispersion equation is expressed in terms of the waveguide cutoff frequency, $\omega_{cq}$.

\begin{align}
k_{zq}=\frac{\sqrt{\omega^2-\omega_{cq}^2}}{c}
\end{align}

Substitution into Eq. \eqref{phase_synchronism} produces a quadratic equation with two solutions, corresponding to two intersections of the beam line with the mode dispersion relation curve of Fig.~\ref{zeroslippageplot}. In conventional FEL configurations, there are always two intersections, and one usually operates at the higher frequency intersection point. For the tangential intersection point satisfying conditions \eqref{phase_synchronism} and \eqref{vel_res}, the two solutions of the quadratic equation must be degenerate. This results in \cite{gainregimeJerby}

\begin{align}
\omega_0=\gamma_z\omega_{cq}=\gamma_z^2 \beta_z ck_u
\end{align}
where
\begin{align}
\gamma_z=\lr{1-\beta_z^2}^{-\frac{1}{2}}=\frac{\gamma}{\sqrt{1+K^2/2}}
\end{align}

Here $\beta_z=v_z/c$ is the average axial velocity of the beam, and $K=\frac{eB_u}{k_u mc}$ is the undulator parameter, where $B_u$ is the magnetic field amplitude of a periodic linear undulator. Note that in the relativistic limit,
\begin{align}
\omega_0=\gamma_z^2 ck_u
\end{align}
is exactly half the frequency of the FEL synchronism frequency in free space propagation.

To summarize, in order to maximize the efficiency of the energy exchange between the near single-cycle THz wave and the electron beam, it will be necessary to choose the undulator and waveguide parameters to meet the "zero-slippage" condition (satisfying both Eq. \eqref{vel_res} and Eq. \eqref{phase_synchronism}) which will keep the beam in contact with the maximum available driving field for the entire length of the interaction.

As the broad spectrum, near single-cycle THz pulse undergoes dispersion within the guide, the amplitude envelope may become less sharply peaked, relaxing this velocity-matching condition. This effect needs to be carefully considered when looking at the frequency-dependent radiation pulse evolution in the waveguide.

To be more quantitative, we start by assuming the incident field to be well matched to the waveguide so that only a single mode is excited. We can then write the field as a Fourier integral in order to explicitly account for the disparate evolution of the spectral components.
\begin{align}\label{E_field_init}
\vc{E}(\vc{r},t) = \frac{1}{2\pi} \int_{-\infty}^{\infty} c_q(z,\omega ) \vep_q (x,y,\omega ) e^{i k_{zq}(\omega )z - i \omega t} d\omega
\end{align}
$\vep_q(x,y,\omega)$ is the transverse profile of mode $q$ and the $c_q(z,\omega)$ coefficients are determined by the Fourier transform of the field at the entrance, $z=0$. For a smooth waveguide, such as the parallel plates with which we will be working, we may assume the transverse profile is dispersionless, eliminating the frequency dependence in $\vep_q (x,y,\omega)$. Because $\vc{E}(\vc{r},t)$ is real, the expression then simplifies to

\begin{align}\label{E_field}
\vc{E}(\vc{r},t) = \frac{1}{\pi} \text{Re}\Bigg{[}\vep_q (x,y)\int_{0}^{\infty} c_q(0,\omega ) e^{i k_{zq}(\omega )z - i \omega t} d\omega \Bigg{]}
\end{align}

Using the expanded expression for the propagation wavenumber around the peak frequency of our source,
\begin{align}
k_{zq}(\omega) = k_{zq}(\omega_0) + \frac{d k_{zq}}{d \omega}\big{|}_{\omega_0} (\omega-\omega_0) + \Delta k_z(\omega,\omega_0)
\end{align}
where $\frac{d k_{zq}}{d \omega}\big{|}_{\omega_0}$ is defined to be the inverse of the group velocity, $1/v_g$, we can gain insight into the evolution of the pulse envelope. $\Delta k_z(\omega,\omega_0)$ represents terms of second order and higher in $\omega-\omega_0$. Inserting this expansion into Eq.~\eqref{E_field} and factoring the $\omega$-independent terms out of the integrand, we have

\begin{align}
\vc{E}(\vc{r},t) ~=~\frac{1}{\pi} \text{Re}\Bigg{[}\vep_q(x,y) \underbrace{e^{i k_{zq}(\omega_0 )z - i \omega_0 t}}_\text{carrier wave} \underbrace{\int_{0}^{\infty} c_q(0,\omega) e^{-i (\omega-\omega_0)\lr{t-\frac{z}{v_g}} + i \Delta k_z(\omega,\omega_0) z} d\omega}_\text{time-dependent envelope} \Bigg{]} \text{.}
\end{align}

We can interpret the form of this description of the THz pulse in terms of a sinusoidal carrier wave that accumulates phase $k_{z0}z$, propagates with constant phase velocity $\omega_0/k_{z_q}(\omega_0)$, and is modulated by a time-dependent envelope. Only the higher order terms in the envelope function are time dependent, as $t-z/v_g$ reduces to an initial time constant. When the higher order terms are negligible, the envelope function becomes time-independent, indicating that the pulse propagates without distortion. We will refer to this envelope function as $f_{\env}(t-z/v_g)$ in our discussion of a THz streaking application.

\begin{figure}[!htb]
   \centering
   \includegraphics*[scale=.35]{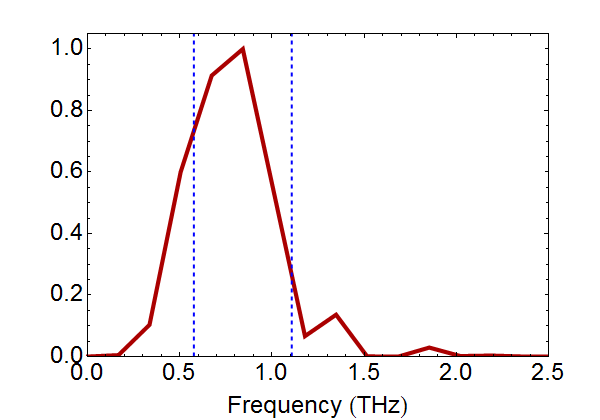}
   \caption{Spectrum of the THz pulse from EOS measurements with the limits of the distortionless propagation regime marked in blue.}
   \label{frequency_spec}
\end{figure}

For a guide of length $L$, we can estimate a range of frequencies in which the higher orders of the phase expansion can be neglected by enforcing the condition $\Big{|}\Delta k_z(\omega,\omega_0) L \Big{|} \ll \pi$. Using the assumption that only the second order term in the Taylor expansion makes a significant contribution, the range of frequencies that satisfies the condition is
\begin{align}\label{freq_range}
\big{|}\omega -\omega_0 \big{|}  \ll \sqrt{\frac{2\pi}{D L}}
\end{align}

where D is the curvature parameter, $D = \frac{d^2 k_{z_q}}{d \omega ^2}\Big{|}_{\omega_0}$. For the waveguide used to produce the tunable group-velocity results presented below, the frequency limit for distortionless propagation is $0.27 \un{THz}$ from the peak frequency. The spectrum of the THz pulse produced at UCLA includes a wider range of frequencies, as shown in Fig.~\ref{frequency_spec}, requiring numerical integration to track the dispersion of the higher order terms in the time-dependent envelope description.

From this calculation, we see that waveguide-moderated beam manipulation offers not only the enhanced efficiency of a zero-slippage interaction, but also the potential for distortionless propagation for frequencies within the range defined by Eq.~\eqref{freq_range}. With distortionless propagation, the available chirping gradient for beam manipulation is not attenuated by dispersive pulse broadening, thereby extending the interaction length over which strong coupling is possible.

\section*{Phase space manipulation}

In principle, the previous considerations could apply to the design of a THz-based accelerator. Nevertheless, the efficiency of THz power generation is still relatively low (on the order of a percent in the non-linear downconversion process from the laser) and further hindered by low (also below a few percent) wall-plug efficiency of short pulse laser systems. Multiple groups are working towards optimizing the optical rectification process in order to increase the THz wall-plug efficiency by a few orders of magnitude and promote THz as a viable source for particle accelerators \cite{improvingORFulop,coolingHuang,THzgenDASTVicario}. On the other hand, state-of-the-art laser-driven THz sources already have a lot to offer to accelerator science wherever efficiency is not a concern, as with beam phase space manipulations such as longitudinal compression, energy dechirping, and, as we will discuss in the last section of the paper, transverse streaking.

A THz IFEL offers the possibility of total capture of the electron beam within a single ponderomotive bucket which can result in compression ratios exceeding 10x. For optimum compression, the electron beam should be injected in the interaction at the zero crossing of the THz waveform, so that it will receive a nearly linear energy chirp which a subsequent drift can transform into strong bunching. In practice, the relative phase of the electron bunches will be distributed throughout the ponderomotive bucket because of the inherent timing jitter at the injection. Within the bucket, all electrons with negative phases will increase in energy, while all electrons with positive phases will decrease in energy.

\begin{figure}[!htb]
\centering
\begin{minipage}{.5\linewidth}
\centering
\subfloat[]{\label{initialfieldwA}\includegraphics[scale=.5]{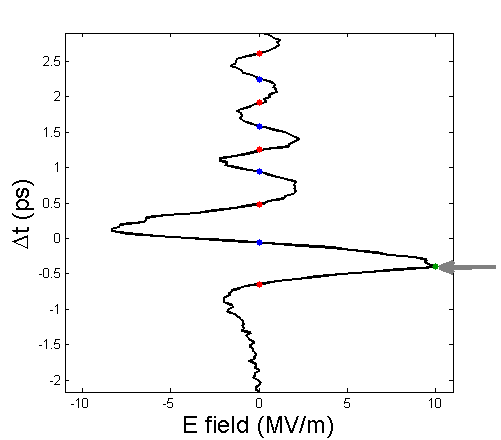}}
\end{minipage}%
\begin{minipage}{.5\linewidth}
\centering
\subfloat[]{\label{zerocrossingmapwA}\includegraphics[scale=.5]{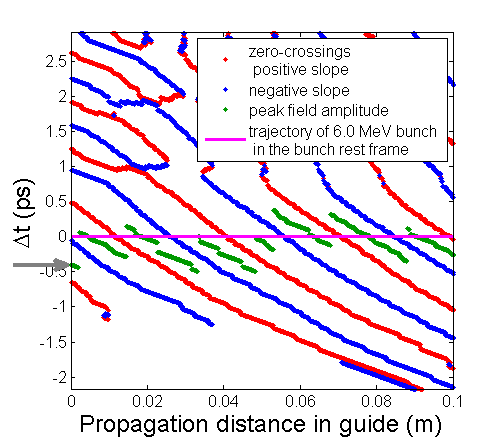}}
\end{minipage}%
\caption{(a) The initial THz pulse waveform with the peak field amplitude marked in green, and the zero crossings marked in red (positive slope) and blue (negative slope). $\Delta t$ on the vertical axis corresponds to the time interval over which the respective point on the THz waveform would propagate before reaching the current longitudinal position of the electron bunch. (b) The zero crossing and peak field markers are plotted, in the rest frame of a $6 \un{MeV}$ electron bunch, as a function of longitudinal position as the pulse propagates inside the first $10 \un{cm}$ of the guide. The trajectory of the electron bunch is indicated by the pink line. An example of the correspondence between the initial snapshot at left and the beginning of the marker evolution at right is indicated by the gray arrows.}
\label{optimizetrajectory}
\end{figure}

To determine the resonant phase for injection of the bunch into the ponderomotive bucket, we can examine the distribution of the zero crossings of the THz waveform as it evolves inside the waveguide. In Fig.~\ref{zerocrossingmapwA}, the relative temporal positions, $\Delta t$, of each zero crossing within the THz pulse profile are plotted as a function of longitudinal position as the waveform propagates within the waveguide. $\Delta t$ is calculated in the rest frame of a $6\,\un{MeV}$ bunch, so that the trajectory of the bunch, indicated by the pink line, appears stationary in time, while the zero crossings slip past. This plot gives a visual description of what it means to satisfy the phase resonance condition and the zero-slippage condition. The intersections between the trajectory of the electron bunch and the zero crossings of the pulse give the periodicity of the resonant interaction. The green markers show the time stamp of the peak field magnitude; an electron trajectory that stays within the range of the peak field maintains group velocity matching. For optimal performance of the IFEL interaction, we choose an injection phase for the electron bunch that keeps the trajectory close to the peak field.

\begin{figure}[!htb]
\centering
\begin{minipage}{.5\linewidth}
\centering
\subfloat[]{\label{PSbefore}\includegraphics[scale=.4]{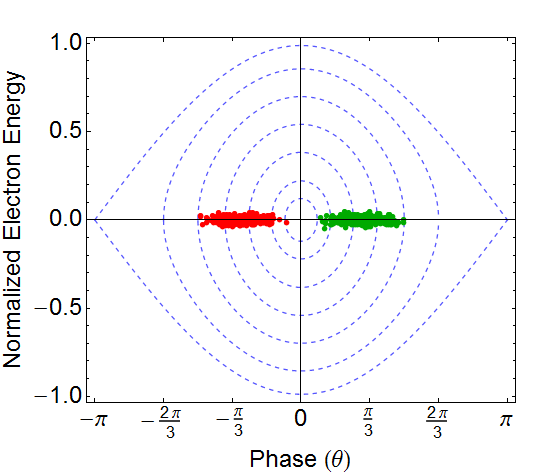}}
\end{minipage}%
\begin{minipage}{.5\linewidth}
\centering
\subfloat[]{\label{PSafter}\includegraphics[scale=.4]{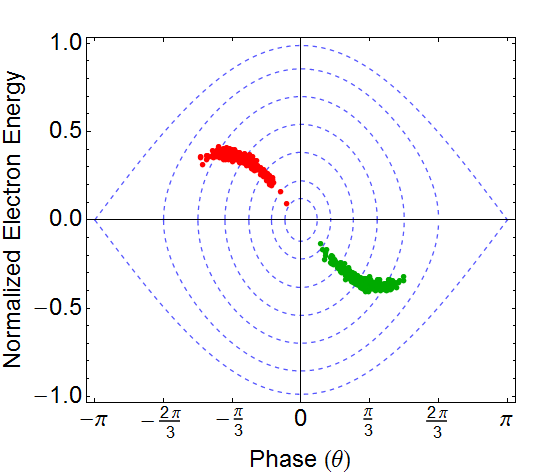}}
\end{minipage}%
\caption{Phase space plots showing (a) the initial distributions of an "early" (green) and "late" (red) electron bunch within the ponderomotive bucket, indicated in blue, and (b) their chirped phase space distributions after evolving under the ponderomotive force. The normalized electron energy is defined by $\frac{\gamma-\gamma_0}{\gamma_0}$, where $\gamma_0$ is the mean Lorentz factor of the bunch.}
\label{phase_space_rotation}
\end{figure}

Two snapshots in the evolution of this process for a "late" (red) and "early" (green) bunch can be seen in Fig.~\ref{phase_space_rotation}. With the increase in energy, the "late" electrons will travel faster, catching up towards the equilibrium phase over a drift section. Similarly, "early" electrons will travel slower, slipping back towards the equilibrium phase over a drift section. An electron bunch arriving within $\pi/2$ of the equilibrium phase will experience an energy chirp that accelerates the particles at the back towards the front of the bunch or decelerates the particles at the front towards the back of the bunch. Farther from the resonant phase the energy chirp becomes less linear. If the phase space distribution of a bunch is far from the resonant phase, it will fall out of the longitudinal acceptance of the compressor and the beam will acquire energy spread without decreasing the bunch length.

Fluctuations of the input energy from the resonant condition would also displace the bunch from optimal positioning within the ponderomotive bucket. Phase space rotation of a bunch with a small energy deviation would result in only slightly less effective bunching and a slight increase in the mean energy offset. In practice however, the arrival time of the electrons would also be affected by the electron beam energy. The phase space rotation within the ponderomotive bucket may then work to reduce the timing jitter in this case, but it can also over-correct the mean energy of the bunch, increasing the final energy deviation.

\section*{THz waveguide}

Although higher frequencies can offer larger acceleration gradients, the size of the guiding structures necessary for a waveguide-controlled interaction becomes prohibitively small for co-propagation of electrons and laser. The length scales necessary for a THz guiding structure are large enough to permit alignment of the electron beam without clipping. For reasonably low charge, wakefield effects in the guiding structure are negligible. The guided propagation has the added benefit of preserving the field intensity over the length of the guide rather than operating in the diffraction limited regime where high intensity would be particularly difficult to maintain for THz frequencies.

\begin{figure}[!htb]
   \centering
   \includegraphics*[scale=.25]{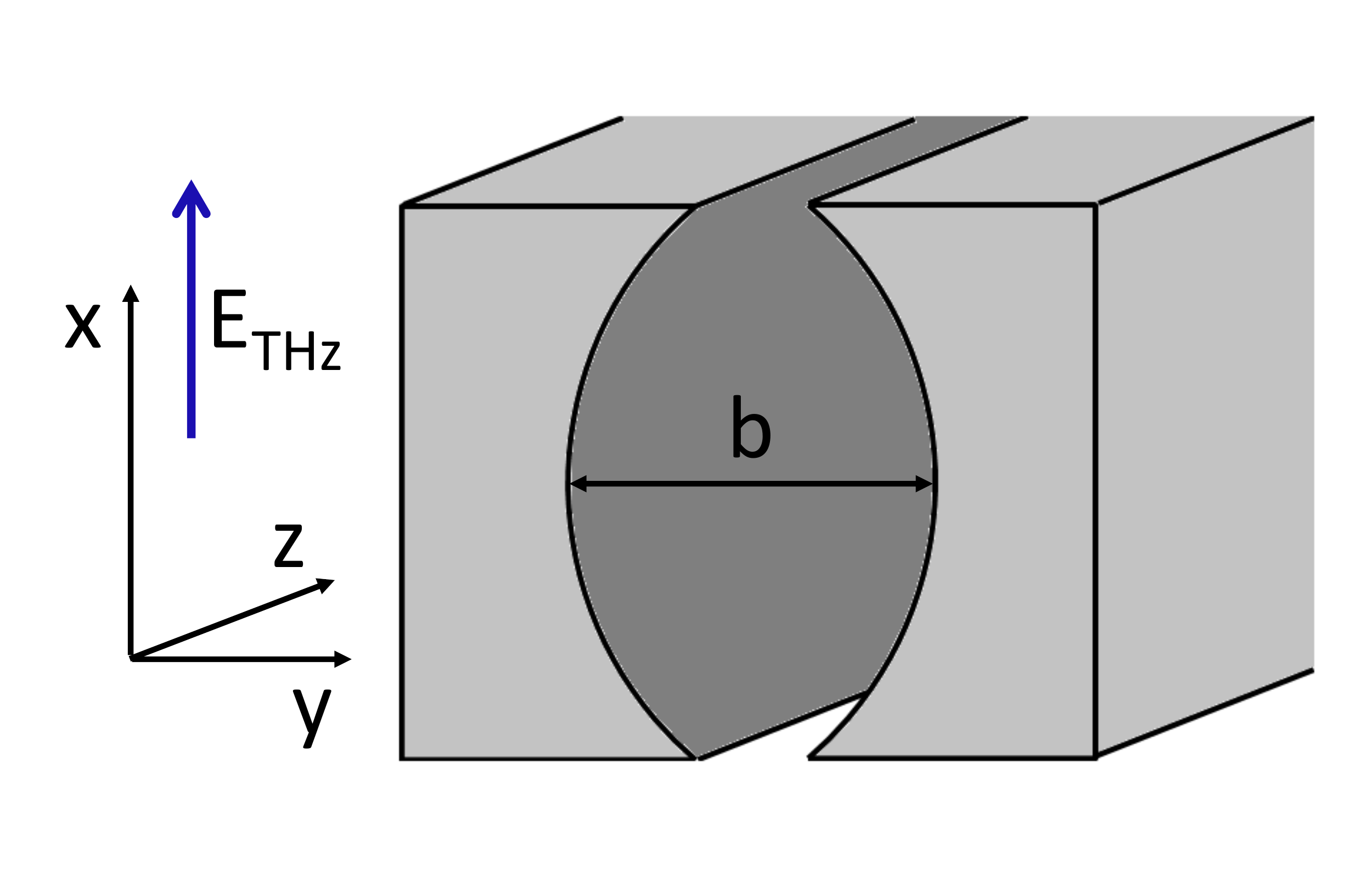}
   \caption{Diagram showing cross section of CPPWG, with plate spacing b, and the orientation of THz polarization to excite the $\text{TE}_{01}$ mode.}
   \label{PPWG_diagram}
\end{figure}

\begin{figure}[!htb]
\centering
\begin{minipage}{.5\linewidth}
\centering
\subfloat[]{\label{atten_diffr}\includegraphics[scale=.4]{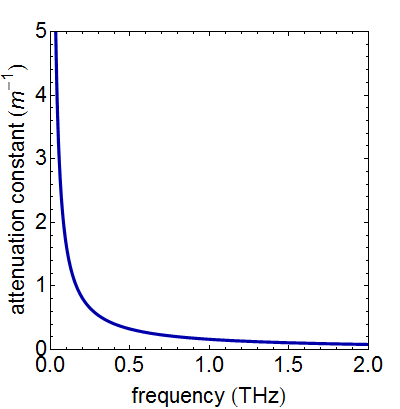}}
\end{minipage}%
\begin{minipage}{.5\linewidth}
\centering
\subfloat[]{\label{freq_cutoff}\includegraphics[scale=.4]{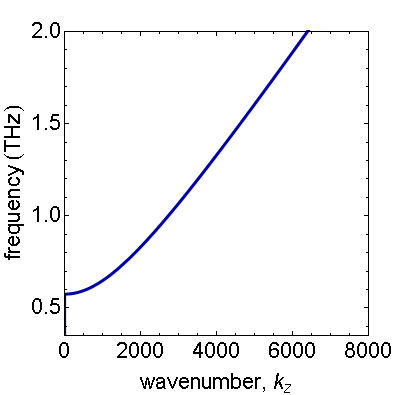}}
\end{minipage}%
\caption{Plots of (a) the attenuation due to diffraction and (b) the dispersion relation for the waveguide parameters, shown in Table \protect{\ref{parameters}}, that are used in simulations of the guided-THz IFEL interaction.}
\label{PPWG_properties}
\end{figure}

We have adopted the curved parallel plate waveguide (CPPWG), shown in Fig.~\ref{PPWG_diagram}, to control the propagation of the THz pulse. This choice was motivated by the investigation of guiding structures for a THz FEL oscillator in ref.~\cite{waveguidestudyYakover} and~\cite{resonatordesignYakover}. The parallel plate structure has small ohmic losses relative to more conventional waveguides and offers the unique advantage of variable plate spacing. This dynamic control of the waveguide parameters allows for tuning of the guide's dispersive properties to optimize the group velocity of the THz pulse for our application. The gap between plates results in some attenuation due to diffraction, but this effect is mitigated by adding a curvature to the plates which focuses the THz within the waveguide channel. Attenuation due to diffraction diminishes with increasing frequency, as shown in Fig.~\ref{atten_diffr}. This feature results in a decrease in total attenuation with increasing frequency, unlike conventional waveguides. While decreasing the plate spacing reduces the group velocity to match the propagating beam, it also increases the cut-off frequency within the waveguide, see Fig.~\ref{freq_cutoff}, and increases the rate of pulse broadening within the guide. The utility of the THz pulse is eventually limited by the dispersive pulse broadening that reduces the available field gradient for chirping the beam as the pulse propagates down the length of the guide.

The THz mode profile in the CPPWG is determined by the dispersion relation giving the transverse wavenumber, $k_{mn}$, and longitudinal wavenumber, $k_z$, 

\begin{align}
\begin{split}
k_{mn} &= \frac{1}{b} \lr{n\pi+(2m+1)\tan^{-1}\frac{b}{\sqrt{2Rb-b^2}}} \\
k_z &= \sqrt{k_0^2-k_{mn}^2}\text{ .}
\end{split}
\end{align}

\begin{figure}[!htb]
\centering
	\begin{minipage}{.5\linewidth}
	\centering
	\subfloat[]{\label{TE01_in}\includegraphics[scale=.4]{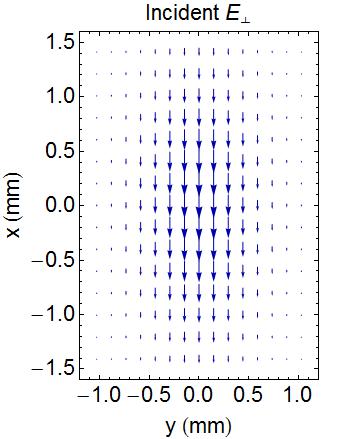}}
	\end{minipage}%
	\begin{minipage}{.5\linewidth}
    \centering
    \subfloat[]{\label{TE01_out}\includegraphics[scale=.4]{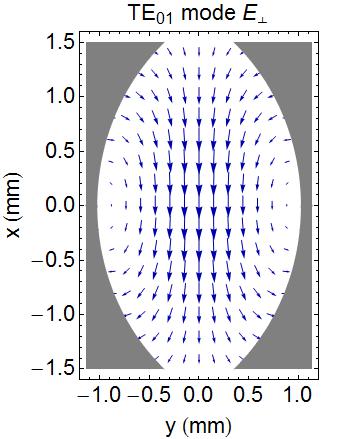}}
	\end{minipage}%
\caption{Plots of (a) the incoming THz field profile matched to (b) the transverse profile of the $\text{TE}_{01}$ mode within the CPPWG cross section. The CPPWG plates are shown in gray.}
\label{TE01_mode}
\end{figure}

The form of the longitudinal field, $\Phi_{mn}$, corresponding to $H_z$ for TE modes and $E_z$ for TM modes, is given in terms of the Hermite polynomials, $\He_{\, n}$, as follows \cite{ppwgNakahara}

\begin{align}
\Phi_{mn} = \frac{e^{\frac{-\beta_{mn}^2 x^2}{\alpha_{mn}(y)}}}{\sqrt[4]{\alpha_{mn}(y)}} \, \He_{\, n} \lr{\frac{2\beta_{mn}x}{\sqrt{\alpha_{mn}(y)}}} \lr{
\begin{matrix}
\renewcommand*{\arraystretch}{2}
 \cos \\ \sin
\end{matrix}}
\Bigg[   k_{mn}y+\frac{2\beta_{mn}^4yx^2}{k_{mn}\alpha_{mn}(y)}-\lr{m+\frac{1}{2}} \tan^{-1}\frac{2 \beta_{mn}^2y}{k_{mn}}   \Bigg] e^{\pm i k_z z}
\end{align}
where 
\begin{align}
\begin{split}
\alpha_{mn}(y) &= 1+4\frac{\beta_{mn}^4 y^2}{k_{mn}^2} \\ 
\beta_{mn} &= \sqrt{\frac{k_{mn}}{\sqrt{2Rb-b^2}}} 
\end{split}
\end{align}
and $R$ and $b$ are the plate curvature and spacing of the CPPWG. The transverse E-field components are then given by

\begin{align}
\begin{split}
&E_x = \frac{-i}{k_{mn}^2} \lr{k_z \frac{\partial E_z}{\partial x}+\omega \mu_0 \frac{\partial H_z}{\partial y}} \\
&E_y = \frac{-i}{k_{mn}^2} \lr{k_z \frac{\partial E_z}{\partial y}-\omega \mu_0 \frac{\partial H_z}{\partial x}}
\end{split}
\end{align}
The $\text{TE}_{01}$ mode of the curved CPPWG is optimal for ponderomotive coupling to the beam. The transverse profile, shown in Fig.~\ref{TE01_mode}, has a peak field on-axis with polarization parallel to the long axis of the CPPWG cross-section. This orientation allows maximum clearance for the parallel wiggle of the electron trajectory within the guide. For the simulation parameters listed in Table~\ref{parameters}, the maximum amplitude of the oscillating electron bunch trajectory is $457 \text{\,}\mu \text{m}$. At this distance, the THz field magnitude has decreased by only $6.5\%$ from the on-axis maximum. When the E-field is linearly polarized in the direction parallel to the plate surfaces, direct coupling of a THz pulse with a bi-Gaussian profile, as shown in Fig.~\ref{TE01_in}, excites the $\text{TE}_{01}$ mode of the CPPWG with $82\%$ efficiency.

\begin{figure}[!htb]
   \centering
   \includegraphics*[scale=.7]{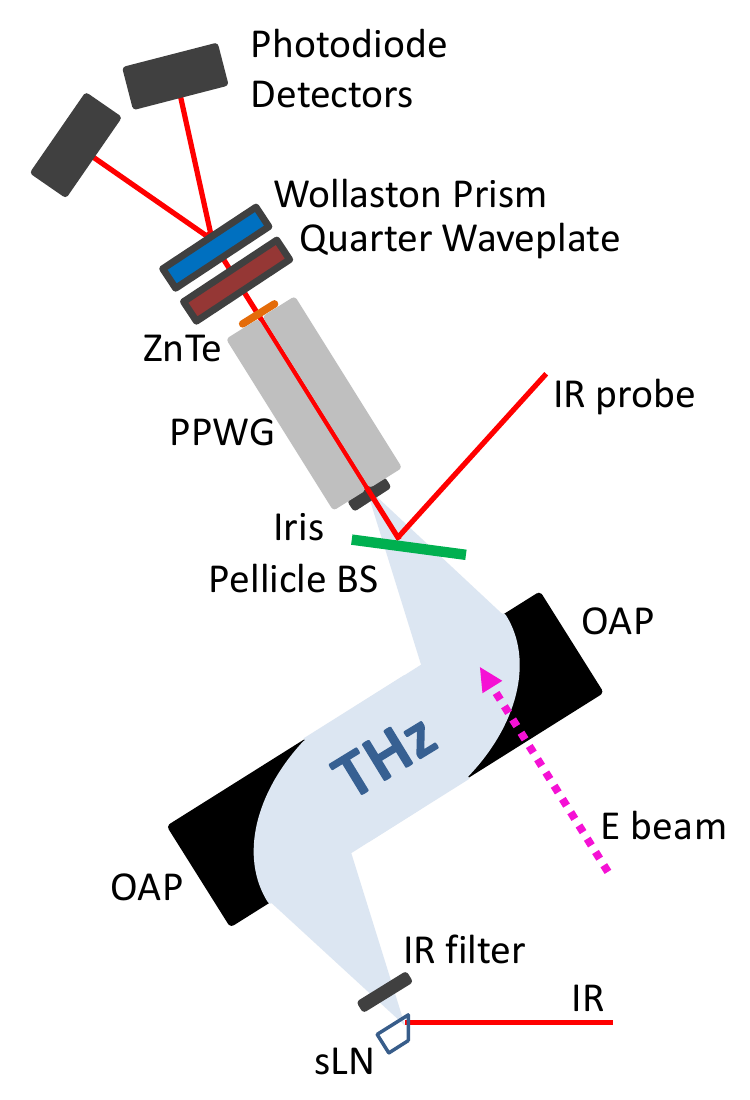}
   \caption{Diagram showing the path of the THz pulse as it is collimated then focused into the CPPWG and the path of the IR for the EOS measurement. The planned electron trajectory is shown in pink.}
   \label{set_up_diagram}
\end{figure}

\begin{figure*}[!htb]
\centering
\begin{minipage}{.9\linewidth}
\centering
\par\subfloat[]{\label{sub1d5m}\includegraphics[scale=.4]{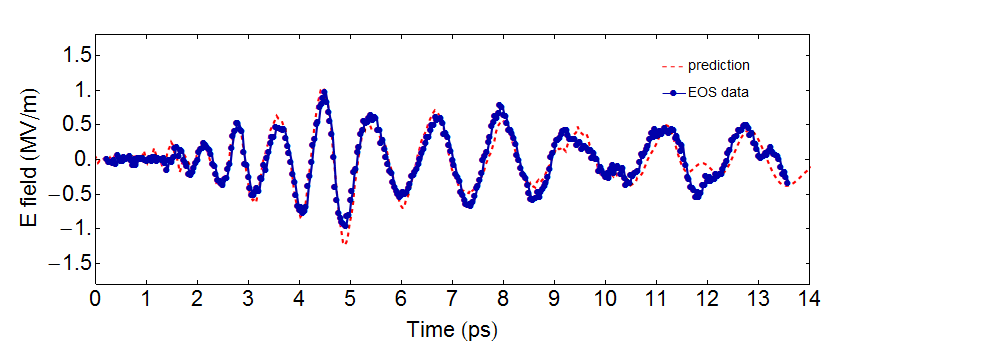}}
\par\subfloat[]{\label{sub2d5m}\includegraphics[scale=.4]{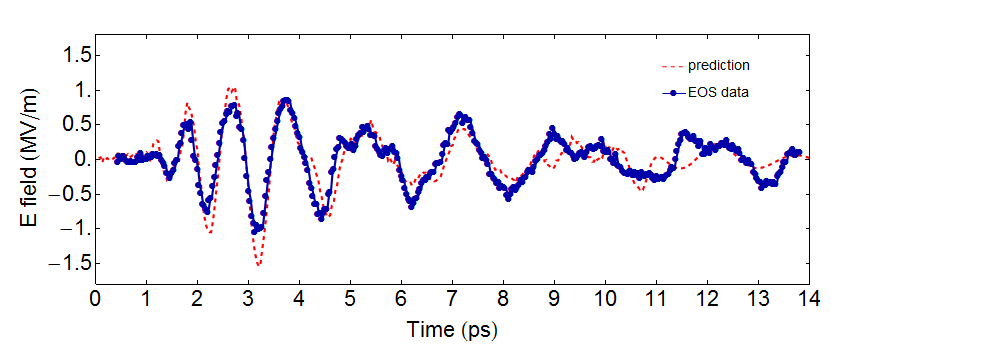}}
\par\subfloat[]{\label{sub3d5m}\includegraphics[scale=.4]{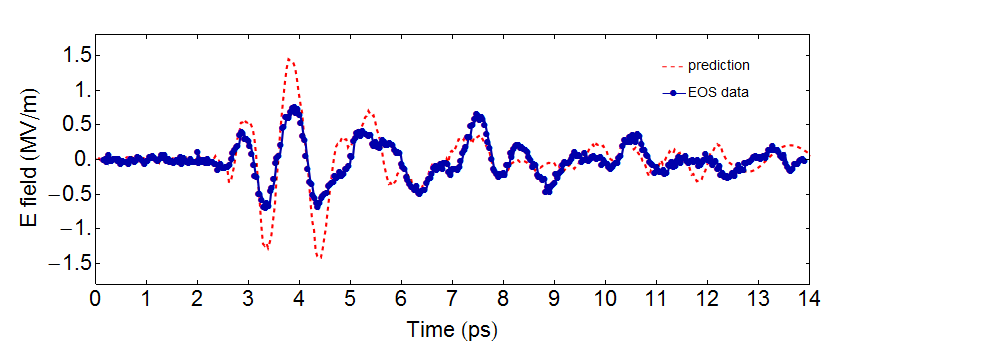}}
\end{minipage}%
\caption{EOS measurements of the THz profile after a $15\un{cm}$ CPPWG with plate spacings of (a) $1.5\un{mm}$,  (b) $2.5\un{mm}$,  and (c) $3.5\un{mm}$. The red dashed line shows the simulated THz profile based on an EOS measurement of the pulse waveform at the entrance of the CPPWG, see Fig.~\protect{\ref{UDnoPPWG}}. The timing offset of the simulated pulse has been adjusted to overlap with EOS measurements.}
\label{PPWGprop}
\end{figure*}

\section*{Guided THz propagation}

Measurements of the temporal field profile of the THz pulse produced at PEGASUS laboratory are conducted using electro-optic sampling (EOS). An IR probe pulse is brought into collinear propagation with the THz using a pellicle that is transparent to THz, as shown in Fig.~\ref{set_up_diagram}. We focus the THz pulse into the entrance of the CPPWG using an off-axis parabolic mirror (OAP). When this configuration is adapted for bunch compression, a small hole will be drilled through the center of the OAP to allow electrons to pass through and co-propagate with the THz pulse. Because the spot size of the THz on the surface of the mirror is large relative to the electron window, the power loss will be minimal.

The IR probe is aligned to the THz focus using a pyroelectric detector to ensure spatial overlap within the ZnTe. The CPPWG plates are then aligned to the IR probe line, to ensure coupling of the focused THz pulse into the CPPWG entrance. For measurements of the THz profile after the waveguide, the IR probe travels within the guide, before interacting with the THz in the ZnTe crystal placed at the exit. The IR probe pulse is undistorted in the guide, because its wavelength is much smaller than the waveguide dimensions.

In Fig.~\ref{PPWGprop}, we show the THz longitudinal pulse profile after a $15\un{cm}$ CPPWG for three different plate spacings. The plates, made from aluminum with a $3\un{mm}$ radius of curvature and $4.25\un{mm}$ thickness, were manufactured at UCLA using conventional machining techniques. The predicted pulse profile for each plate spacing, shown as the red dashed line in Fig.~\ref{PPWGprop}, was evolved from an initial pulse waveform determined by EOS measurements at the entrance to the guide and neglects losses due to diffraction. To account for imperfect coupling efficiency, the magnitude of the initial field used for the simulations was scaled to match the $1.5\un{mm}$ spacing measurement. For the $1.5\un{mm}$ plate spacing, where attenuation due to diffraction is minimal, there is good agreement between the measured and predicted waveforms. For wider plate spacings, the increased attenuation due to diffraction results in a greater discrepancy between the measured and predicted field magnitudes and waveforms.

\begin{figure}[!htb]
   \centering
   \includegraphics*[scale=.35]{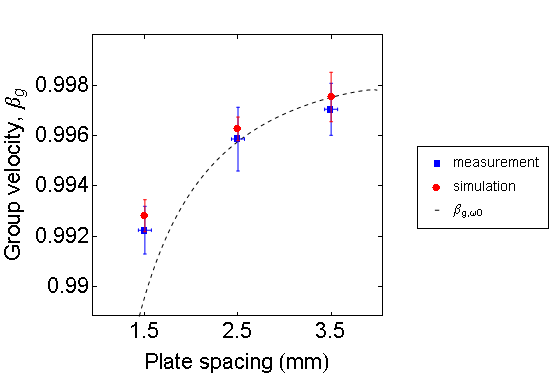}
   \caption{Group velocity estimates for three plate spacings determined from EOS measurements before and after the waveguide (blue squares) and simulated propagation of the initial pulse profile (red circles). The theoretical group velocity at the peak frequency of the pulse, $.84\un{THz}$, is shown by the black dashed line.}
   \label{group_velocity}
\end{figure}

Using EOS measurements of the THz pulse profile at the entrance and exit of the CPPWG, a direct estimate of the group velocity of the pulse within the waveguide can be obtained from the IR probe delay between the peaks of the initial and final pulse envelops. The envelope of the pulse is extracted from the raw data by factoring out the sinusoidal carrier wave determined by a least-squares fit in the region of the peak field. Experimental results for the group velocity determined using this method are shown in Fig.~\ref{group_velocity}, along with the corresponding estimates found via application of this method to the simulated pulse propagation. The theoretical group velocity at the peak frequency of the pulse, shown by the black dashed line in Fig.~\ref{group_velocity}, provides a reasonable estimate for the expected group velocity for the case of $2.5\un{mm}$ and $3.5\un{mm}$ plate spacings. For the $1.5\un{mm}$ plate spacing, both measurement and simulation deviate from the group velocity at the peak frequency. This behavior may be explained by the highly nonlinear dispersion at low frequencies which causes a frequency chirp that reduces the contribution of the low frequency components to the pulse envelop peak. The speed of the envelop peak is then determined by the high frequency components, with their correspondingly high group velocities. For the larger plate spacings, the region of near-linear dispersion extends to lower frequencies, making the effect less significant. 

\section*{Guided-THz IFEL simulations}

\begin{figure*}[!htb]
\centering
\includegraphics*[scale=.2]{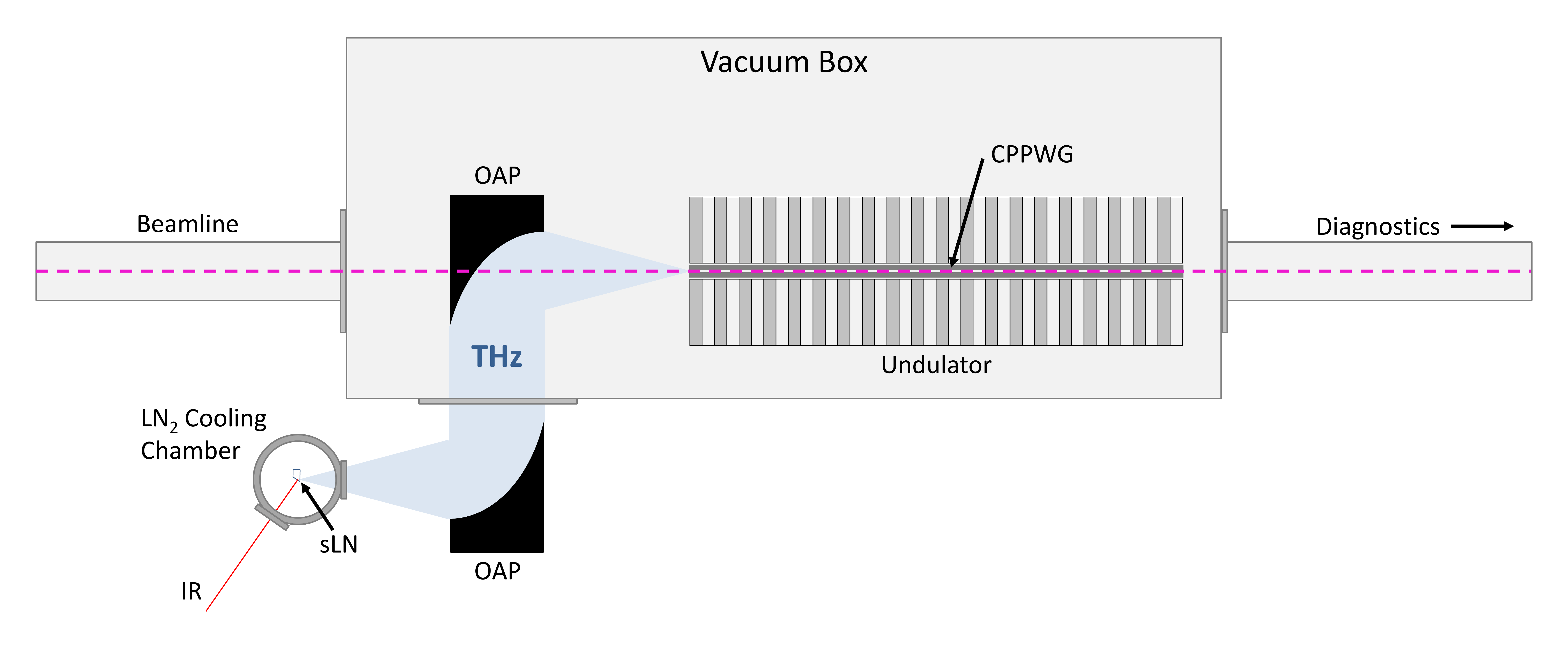}
\caption{Diagram of the guided-THz IFEL compressor set-up. Optical rectification within the sLN crystal, cooled by liquid nitrogen, produces the THz pulse which is collimated into the vacuum box with an OAP. The path of the $e^-$ beam (shown in pink) passes through the OAP to come into co-propagation with the THz before entering the CPPWG nested inside the undulator. The THz source set-up will remain outside of the vacuum box to facilitate independent testing and optimization.}
\label{vacuum_box}
\end{figure*}

\begin{figure}[!htb]
   \centering
   \includegraphics*[scale=.5]{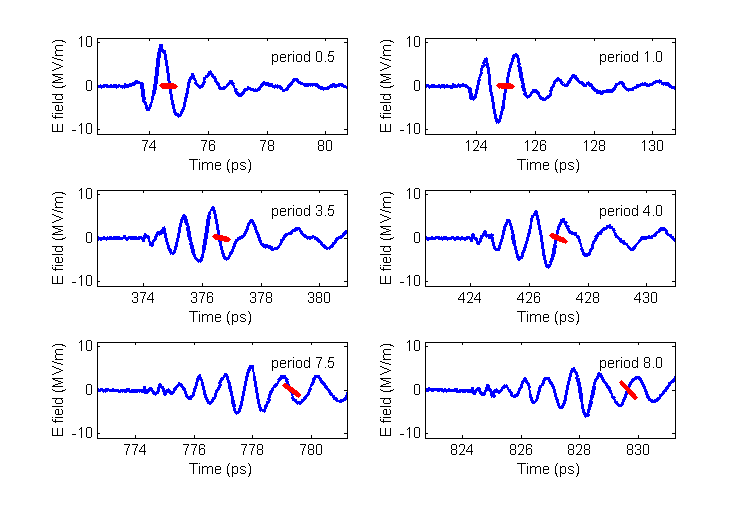}
   \caption{Snapshots of the simulated THz pulse propagation within the waveguide, after $0.5 \lambda_u$, $1 \lambda_u$, $3.5 \lambda_u$, $4 \lambda_u$, $7.5 \lambda_u$, $8 \lambda_u$, with electron bunch position shown in red. As the system is evolved in terms of the longitudinal position coordinate, the time marked on the x-axis indicates the time required for the corresponding point on the THz waveform to cross the current longitudinal coordinate. The plot window tracks with propagation at the speed of light, therefore the THz pulse and the electron bunch appear to slip further back with each snapshot as they propagate at $.994 c$. Note that at the half period intervals the sign of the THz field gradient flips.}
   \label{trackzerocrossing}
\end{figure}

A layout of the guided-THz IFEL compressor is shown in Fig.~\ref{vacuum_box}. The undulator and waveguide parameters used for simulation of the guided-IFEL interaction are shown in Table~\ref{parameters}. The chosen waveguide parameters correspond to a theoretical group velocity of $\beta_g = .994$ at the measured peak frequency of the PEGASUS THz source, within the range that has been experimentally demonstrated using our tunable CPPWG. The simulated electron bunch parameters are well within the range that can be routinely generated at PEGASUS laboratory~\cite{pegasusMusumeci}~\cite{pegasusMoody}. With these constraints, the undulator parameters have been calculated from Eq.~\eqref{phase_synchronism}, the resonance condition modified for interaction within a waveguide, such that the average longitudinal beam velocity matches the THz group velocity. In Fig.~\ref{trackzerocrossing}, we show the co-propagation of the simulated THz pulse and electron beam for this resonant interaction.

\begin{table}[hbt]
   \centering
 \caption{Simulation Parameters}
   \begin{tabular}{lcc}
   	       \hline
           Bunch energy             & $6\un{MeV}$              \\
           Bunch charge				& $1\un{pC}$ 			   \\
           Energy spread            & $3\times 10^{-4}$                 \\
           Bunch length             & $100\un{fs\,rms}$        \\
          Undulator period          & $3\un{cm}$             \\
       Undulator parameter, K       & $1.12$         \\
       \# of undulator periods      & $8$                   \\
           CPPWG spacing, b         & $2.06\un{mm}$            \\
         Plate curvature radius     & $2\un{mm}$            \\
           Peak frequency           & $.84\un{THz}$              \\
           Peak THz field           & $10\un{MV}/\un{m}$            \\
       \hline
   \end{tabular}
   \label{parameters}
\end{table}

Ponderomotive coupling with a near single-cycle THz pulse cannot be well-described by the single frequency approximation used by most FEL simulation codes because of the broad spectral content of the THz pulse. We are currently developing a multi-frequency simulation code, similar to MUFFIN \cite{muffinPiovella}, which will track the evolution of the THz pulse, described by Eq.~\eqref{E_field}. The calculation does not rely on the slowly varying envelope approximation or period averaging and includes the waveguide-induced dispersion of the THz pulse.

\begin{figure}[!htb]
\centering
	\begin{minipage}{.3\linewidth}
	\centering
	\subfloat[]{\label{initial}\includegraphics[scale=.4]{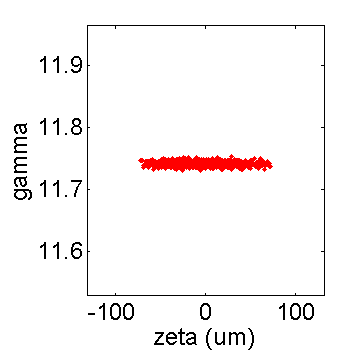}}
	\end{minipage}%
	\begin{minipage}{.3\linewidth}
    \centering
    \subfloat[]{\label{chirped}\includegraphics[scale=.4]{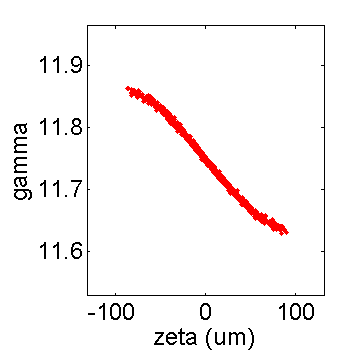}}
	\end{minipage}%
	\begin{minipage}{.3\linewidth}
    \centering
    \subfloat[]{\label{compressed}\includegraphics[scale=.4]{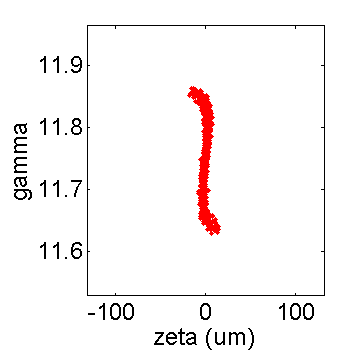}}
	\end{minipage}%
\caption{Simulation results showing the longitudinal phase space distribution (a) before the undulator, (b) at the exit of the undulator, and (c) after a $1.03\un{m}$ drift period.}
\label{waffel_run}
\end{figure}

For very low beam charges one can neglect the beam current term in the radiation field evolution (frozen field approximation). Preliminary results from our multi-frequency code are shown in Fig.~\ref{waffel_run}, using the design parameters in Table~\ref{parameters}. For these results, the THz pulse was modeled by a Gaussian frequency spectrum peaked at $0.84\un{THz}$ with a FWHM of $.75\un{THz}$ and simulated by 101 spectral points. The peak field was $10\un{MV}/\un{m}$, based on predictions for the cooled sLN conversion efficiency of the PEGASUS optical rectification set-up. The sample electron beam with rms bunch length $100\un{fs}$, shown in Fig.~\ref{waffel_run}, is compressed to $12\un{fs}$. 

\begin{figure}[!htb]
\centering
	\begin{minipage}{.3\linewidth}
	\centering
	\subfloat[]{\label{initialtiming}\includegraphics[scale=.4]{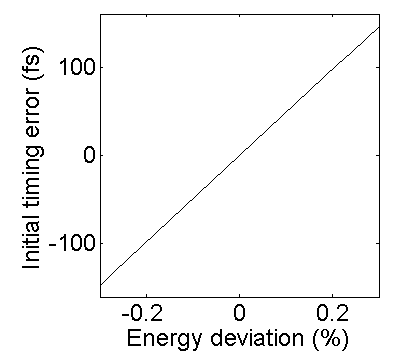}}
	\end{minipage}%
	\begin{minipage}{.3\linewidth}
    \centering
    \subfloat[]{\label{finaltiming}\includegraphics[scale=.4]{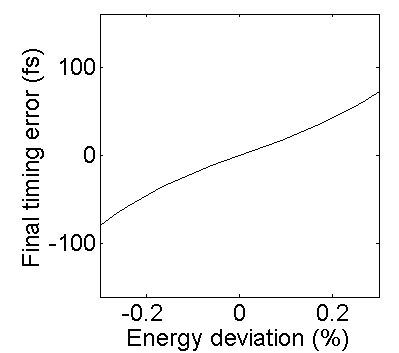}}
	\end{minipage}%
	\begin{minipage}{.3\linewidth}
    \centering
    \subfloat[]{\label{compression}\includegraphics[scale=.4]{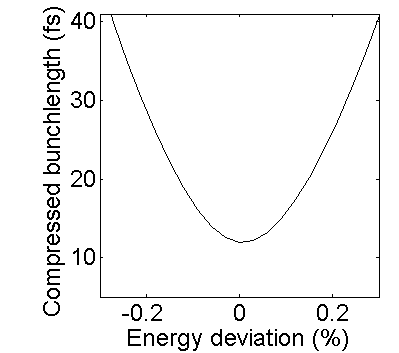}}
	\end{minipage}%
\caption{(a) Initial timing error, (b) final timing error and (c) final bunch compression after varying the mean energy deviation of the initial bunch distribution shown in Fig.~\protect{\ref{waffel_run}}.}
\label{jitter_scan}
\end{figure}

In Figure ~\ref{jitter_scan}, we show the comparative compression and timing jitter of this same bunch distribution for a range of initial mean energy deviations, and the corresponding accrued timing jitter when the bunch enters the IFEL. Even with a relatively large energy deviation of $0.2\%$ and timing error of 100 fs, interaction in the THz-driven IFEL achieves a compression ratio of better than $3$ and cuts the timing jitter by more than half. 

The compression results discussed above are limited primarily by the THz peak field and available space on the PEGASUS beamline. Obviously, increasing the THz field would increase the chirping gradient of the IFEL interaction. Direct improvement of generated THz power is an active area of research. The maximum field attainable via optical rectification in lithium niobate is predicted to be on the order of $100 \un{MV/cm}$, in the range of $0.3-1.5$ THz \cite{improvingORFulop}. Even larger THz fields, on the order of $\un{GV/m}$, have been generated using a partitioned organic DSTMS crystal \cite{THzgenDASTVicario}. Practical improvements can be made by eliminating absorptive media along the THz transport path from the sLN to the waveguide. The size of the vacuum box at the PEGASUS facility prevents cooling the sLN within the chamber, necessitating two transmission windows with intermediate transport through air. However, keeping the sLN outside of the vacuum box does simplify optimization of the alignment for THz generation. The box size places a practical constraint on the number of undulator periods that can fit inside, thereby limiting the duration of the interaction. The pulse frequency spectrum also plays a significant role in constraining the linearity of the energy chirp imparted to the bunch phase space. A longer peak wavelength would allow a larger phase acceptance window and extend the duration over which a $100 \un{fs}$ bunch receives a linear chirp as it rotates in phase space. Unfortunately, while cooling the sLN significantly increases the conversion efficiency of optical rectification, it also results in a slightly blue-shifted THz pulse~\cite{coolingHuang}.  

\section*{THz streak diagnostic}

The THz compressor setup could easily be re-purposed to perform a different THz-driven beam phase space manipulation. With the excitation of an odd symmetric waveguide mode it will be possible to use the THz field to streak the beam, that is, to impart to the electrons a transverse kick correlated with their longitudinal coordinate. After a drift section, the transverse kick maps the longitudinal profile of the beam to the transverse plane, which can then be imaged using a standard fluorescent screen. The use of an odd mode interaction within an undulator was first proposed by Zholents and Zolotorev \cite{streakingZholents} and recently discussed by Andonian et al.~ref. \cite{streakingAndonian}. Preliminary experimental results have confirmed the interaction using a $\text{CO}_{2}$ laser. The wavelength of the $\text{CO}_{2}$ laser is generally small relative to the length of the electron beam, resulting in an alternating transverse kick along the length of the beam. If left to drift unaltered, this snake-like pattern would project onto itself, destroying the imprinted longitudinal information at the diagnostic screen. To discern the different streaks along the length of the beam, ref. \cite{streakingAndonian} incorporates a transverse kick in the orthogonal direction using an RF deflector. The longer wavelength of THz radiation would drastically increase the acceptance of the individual transverse streaks, long enough to capture the entire bunch in one streak when its length is shorter than half a wavelength. 

To calculate the transverse deflection imparted by a guided-THz IFEL interaction in a linear undulator with magnetostatic field in the y-direction, we begin from the force equation
\begin{align}\label{force_eqn}
\frac{d\vc{p}_\perp}{dt} = -e \Big{(}\vc{E}_\perp(\vc{r}_e(t),t) +v_{z0}\hat{\vc{e}}_z \times \vc{B}_\perp(\vc{r}_e(t),t)+\vc{v}_\perp \times \hat{\vc{e}}_z \vc{B}_z(\vc{r}_e(t),t)\Big{)}.
\end{align}

The THz fields are described in terms of the carrier wave and  envelope, $f_{\env}(t-z/v_g)$, as
\begin{align}
\begin{split}
\vc{E}(\vc{r},t) &= Re \left[\E _q(x,y) f_{\env}(t-z/v_g) e^{-i\omega_0 t+ik_{zq}z+i \phi_0} \right] \\
\vc{H}(\vc{r},t) &= Re \left[\bm{\mathcal{H}}_q(x,y) f_{\env}(t-z/v_g) e^{-i\omega_0 t+ik_{zq}z+i \phi_0} \right]
\end{split}
\end{align}
The magnetic field profile in the waveguide is related to the electric field profile by
\begin{align}
\hat{\vc{e}}_z \times \bm{\mathcal{B}}_{q\perp} = \frac{-\E_{q \perp}}{c} \, \Bigg{\{}
\begin{matrix}
\renewcommand*{\arraystretch}{2}
k_{zq}/k_0\text{\, ,} &\text{TE mode  } \\ k_0/k_{zq}\text{\, ,} &\text{TM mode.}
\end{matrix}
\end{align}

For the case of deflection within the plane of the wiggling electron trajectory, the resulting expression is
\begin{align}
\frac{d\vc{p}_x}{dt} = -e \alpha \, Re\left[\E_q (\vc{r}_e(t))f_{\env}(t-z/v_g)e^{-i\omega_0 t+ik_{z0}z+i \phi_0} \right]
\end{align}
where
\begin{align}\label{alpha}
\alpha = \Bigg{\{}
\begin{matrix}
\renewcommand*{\arraystretch}{2}
1-\beta_z \frac{k_{z_q}}{k_0}\text{\, ,} & \text{TE mode  } \\ 1-\beta_z \frac{k_0}{k_{z_q}}\text{\, ,} & \text{TM mode.}
\end{matrix}
\end{align}

The velocity matching condition requires $\beta_z \approx \beta_g$, and the group velocity in the waveguide is given by $\frac{k_{zq}}{k_0}$, as follows from the waveguide property $v_g v_{ph} = c^2$. Eq.~\eqref{alpha} then evaluates to
\begin{align}
\alpha = \Big{\{}
\begin{matrix}
\renewcommand*{\arraystretch}{2}
\frac{1}{\gamma_z^2}\text{\, ,} & \text{TE mode} \\ 0\text{\, ,} & \text{TM mode.}
\end{matrix}
\end{align}
We see that, when sustaining a zero-slippage interaction for deflection in the x-direction, a TE mode is required for the mode-dependent parameter $\alpha$. 

For the antisymmetric mode profile necessary for transverse deflection in the x-direction, we can approximate the profile near the axis as
\begin{align}\label{lin_field}
\E_q(x_e,y_e) \approx \hat{\vc{e}}_x \frac{\partial \mathcal{E}_{qx}}{\partial x} \Big{|}_0 \, x_e(t) = -\hat{\vc{e}}_x \mathcal{E}'_{qx} \, \frac{i K e^{-i k_u z}}{\beta_z \gamma k_u}.
\end{align}
Taking only the phase-synchronous term, the change in $p_x$ becomes
\begin{align}
\frac{d p_x}{dt} =  \frac{e \alpha K}{2 \beta_z \gamma k_u} \, Re \Bigg[ i\mathcal{E}'_{qx} f_{\env}(t-z/v_g) e^{-i\omega_0 t+ik_{zq}z-i k_u z+i \phi_0} \Bigg].
\end{align}
Substituting $z=v_z (t-t_0)$ for an electron entering the wiggler on-axis with velocity $v_z$ at initial time $t_0$, we have
\begin{align}
p_x(t_0,t_r) = \frac{e\alpha K}{2 \beta_z \gamma k_u} \, Re\Big{[}\mathcal{E}'_{qx} e^{-i(\omega_0 t_0-\phi_0-\pi/2)} \int_{0}^{L/v_{z}} f_{\env}\lr{\lr{1-\frac{v_{z}}{v_g}}t+t_0}e^{i((k_{z0}+k_u)v_{z}-\omega_0)t} dt \Big{]}
\end{align}
for an interaction of length $L$. In the zero-slippage case, where $v_{z} \approx v_g$, the resulting angular deflection is
\begin{align}\label{xdeflection}
\Delta \Theta_x (t_0) = \frac{p_x(t_0,L/v_{z})}{m c \gamma \beta_z} = \frac{e\alpha KL \, \mathcal{E}'_{qx} f_{\env}(t_0)}{2 \beta_z^3 \gamma^2 mc^2 k_u} \sinc{\tfrac{\theta L}{2}} \sin{\lr{\omega_0 t_0-\phi_0+\frac{\theta L}{2}}}
\end{align}
where $\theta = \frac{\omega_0}{v_{z}}-(k_0+k_u)$ which is very similar to \cite{streakingZholents}.

By a similar method we can calculate the deflection in the plane perpendicular to the wiggling electron trajectory from the longitudinal coupling in the third term of Eq.~\eqref{force_eqn}. The resulting angular deflection
\begin{align}\label{ydeflection}
\Delta \Theta_y (t_0) = \frac{e KL \, \mathcal{B}_{qz} f_{\env}(t_0)}{2 \beta_z^2 \gamma^2 mc} \sinc{\tfrac{\theta L}{2}} \sin{\lr{\omega_0 t_0-\phi_0+\frac{\theta L}{2}}}
\end{align} 
is optimized for an on-axis B-field.

From both Eq. \eqref{xdeflection} and Eq. \eqref{ydeflection} we can see that, for a streaking application, the angular spread is maximized when $\theta=0$, which is the condition for phase synchronism, and $\omega_0 t_0-\phi_0 = 2\pi m$ for integer $m$, which indicates that the particles are injected at the proper phase in the waveform. Around this point a short bunch of duration $\tau \ll \frac{2\pi}{\omega_0}$ will be angularly spread to 

\begin{align}
\phi_{\text{streak }\{ x,y \}} = \Delta\Theta_{\{ x,y \}} \omega \tau
\end{align}

If this spread is large relative to the intrinsic angular spread in the electron beam, viewing the projected beam spot on a viewer screen reveals the temporal (or axial) distribution of the bunch charge.
 
The $\text{TE}_{11}$ mode of the CPPWG provides an ideal antisymmetric mode profile for deflection in the x-direction, while both the $\text{TE}_{20}$ and $\text{TE}_{02}$ modes are excellent candidates for deflection in the y-direction using the longitudinal B-field, peaked on-axis. The $\text{TE}_{11}$ and $\text{TE}_{20}$ modes could be excited by creating a phase shift in the THz field using two mirror halves shifted by 1/4 of the wavelength in a similar way as employed by Andonian et al. in the $\text{CO}_{2}$ streaking experiments~\cite{streakingAndonian}. A specialized THz waveplate could be employed to excite the $\text{TE}_{02}$ mode or improve the coupling efficiency into the $\text{TE}_{11}$ mode. With either of these simple modification techniques, the experimental hardware from the compression project could be readily adapted for the THz streaking application.

\begin{figure}[!htb]
\centering
\begin{minipage}{.5\linewidth}
\par\subfloat[]{\label{TE11modew}\includegraphics[scale=.25]{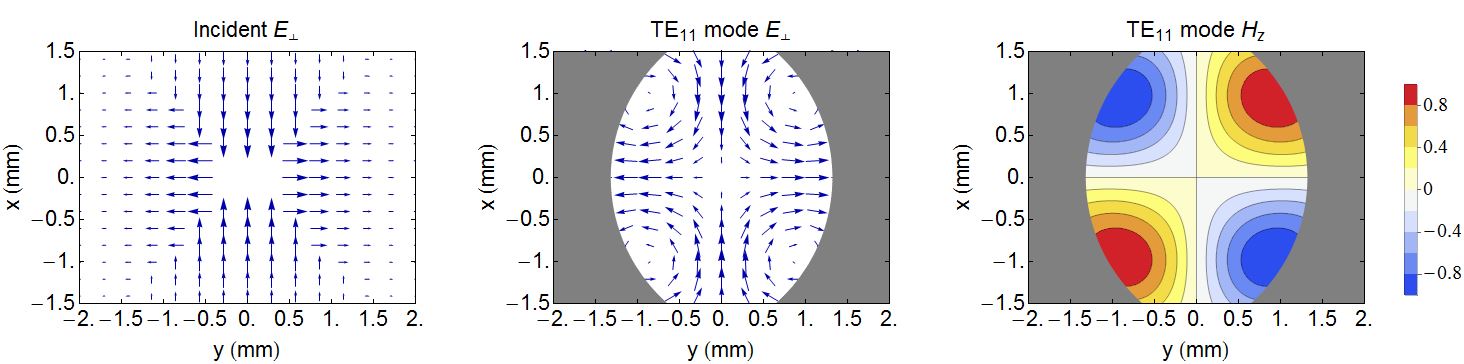}}
\par\subfloat[]{\label{TE11modem}\includegraphics[scale=.25]{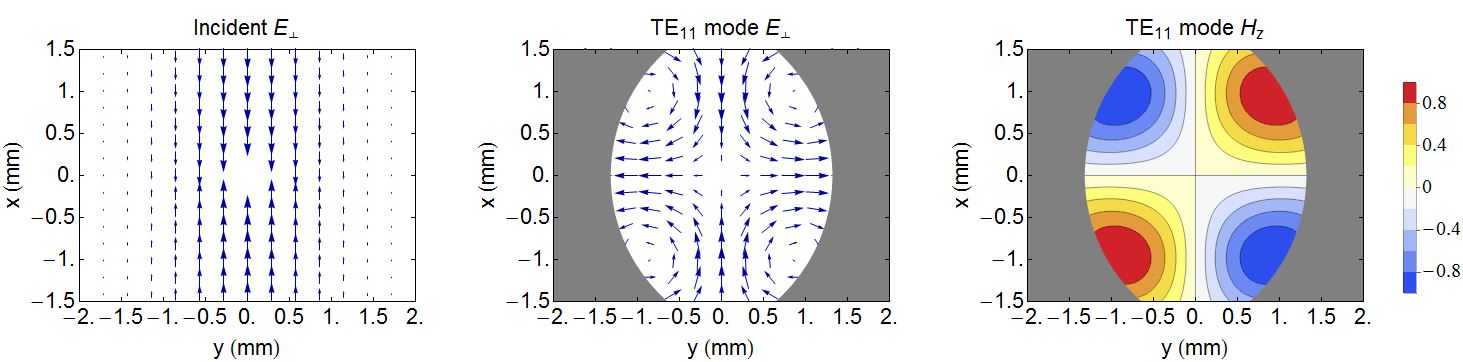}}
\par\subfloat[]{\label{TE02mode}\includegraphics[scale=.25]{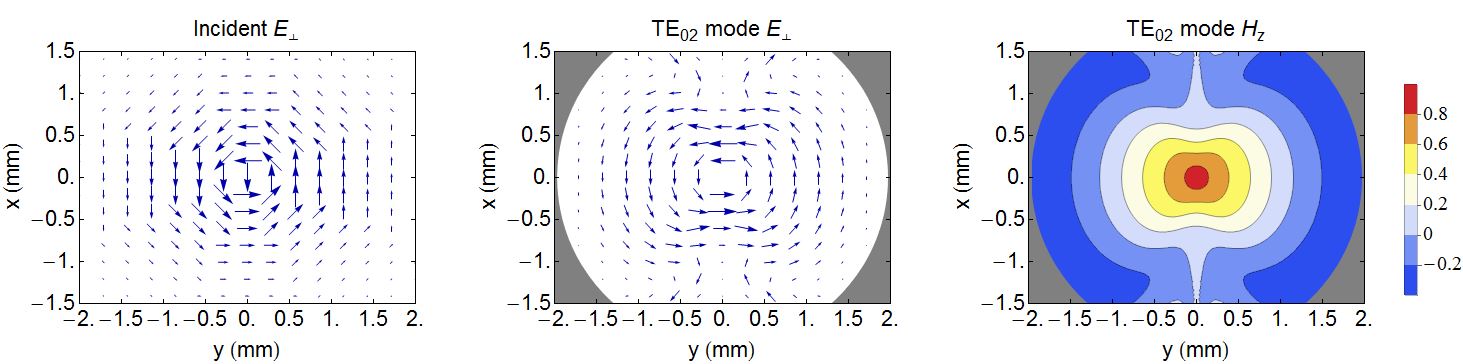}}
\par\subfloat[]{\label{TE20mode}\includegraphics[scale=.25]{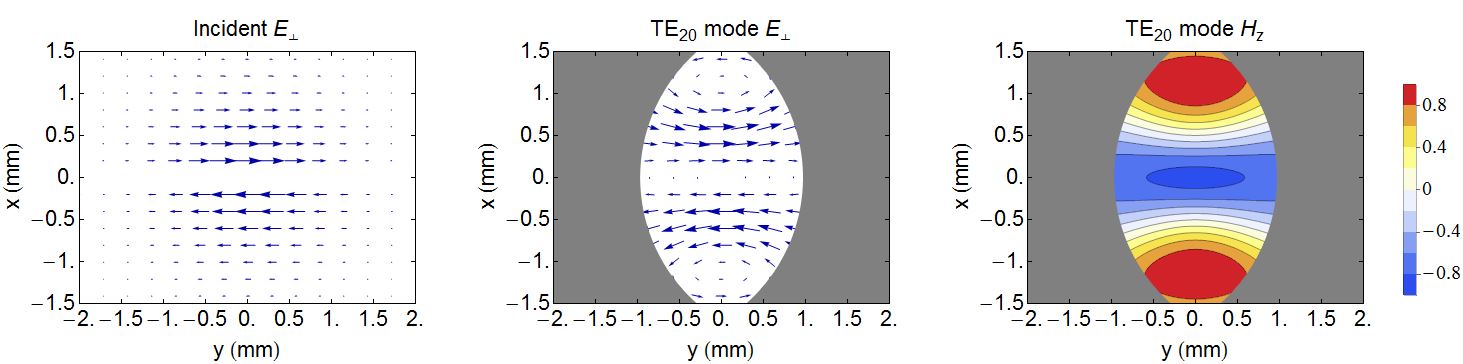}}
\end{minipage}%
\caption{Plots of the transverse mode profile for the (a) $\text{TE}_{11}$ mode, coupled using a waveplate, (b) $\text{TE}_{11}$ mode, coupled using displaced mirror halves, (c) $\text{TE}_{02}$ mode, and (d) $\text{TE}_{20}$ mode, showing, from left to right, the incident THz pulse, electric field of the excited mode, and magnetic field of the excited mode.}
\label{CPPWGprofiles}
\end{figure}

Selection of the optimal streaking method depends on the coupling efficiency of each mode and the corresponding deflection parameter calculated from Eq's \eqref{xdeflection} and \eqref{ydeflection}. A comparison of these values for transverse coupling with the $\text{TE}_{11}$ mode and longitudinal coupling with the $\text{TE}_{20}$ and $\text{TE}_{02}$ mode can be found in Table \ref{streakingcomparison}. These calculations were performed using the undulator parameters fixed by the compression experiment, as shown in Table ~\ref{parameters}. For each mode, the coupling efficiency was calculated over a range of incoming THz profiles that could be produced through a combination of THz waveplate, mirror offset, or non-reflective coating. The incident THz profile offering the maximum coupling efficiency is shown next to the corresponding mode profile in Fig.~\ref{CPPWGprofiles}.

\begin{table*}[!htb]
   \centering
 \caption{THz streaking parameters}
   \begin{tabular}{lcccc}
   	       \hline
									& \textbf{$\text{TE}_{11}$ mode}  & \textbf{$\text{TE}_{11}$ mode}  & \textbf{$\text{TE}_{02}$ mode}  & \textbf{$\text{TE}_{20}$ mode}       \\
       \hline
		   Coupling technique		& waveplate					& offset mirror 		   & waveplate				   & offset mirror  \\
		   Coupling efficiency      & $0.552$					& $0.278$				   & $0.635$				   & $0.638$  \\
           Resonant plate-spacing 	& $2.65\un{mm}$				& $2.65\un{mm}$			   & $3.95\un{mm}$			   & $1.95\un{mm}$  \\
           Resonant energy          & $7.703\un{MeV}$			& $7.703\un{MeV}$		   & $6.148\un{MeV}$		   & $6.225\un{MeV}$   \\
           Angular deflection       & $3.3\un{mrad}$			& $3.3\un{mrad}$		   & $13.9\un{mrad}$		   & $7.9\un{mrad}$  \\
           Effective angular deflection     & $1.8\un{mrad}$	& $0.9\un{mrad}$		   & $8.8\un{mrad}$			   & $5.0\un{mrad}$    \\
           Timing resolution        & $15\un{fs}$				& $30\un{fs}$			   & $4\un{fs}$			       & $7\un{fs}$       \\
           $ \text{   (for }\epsilon_{n,x}=.1\un{mm-mrad} \text{, } \sigma_x=45\text{\,,}\mu\text{m} \text{)}$ \\
       \hline
   \end{tabular}
   \label{streakingcomparison}
\end{table*}

Assuming no position-angle correspondence in the initial beam distribution, the effective timing resolution is defined, for deflection in the x, or y, direction, as \cite{streakingAndonian}
\begin{align}
\Delta t = \frac{\epsilon_{n,\{x,y\}}}{\gamma \sigma_{\{x,y\}} \Delta\Theta_{\{x,y\},\text{max}} c k_z}
\end{align}
where $\epsilon_{n,\{x,y\}}$ is the normalized emittance and $\sigma_{\{x,y\}}$ is the beam size. Despite the low field magnitude from our THz source, we can achieve high temporal resolution, because it scales with the wavelength of the modulating field. Our THz streaking method would be comparable to current femtosecond-scale timing diagnostics, like the X-band transverse deflector at LCLS~\cite{XTCAVresultsBehrens}, with the added advantage of timing synchronization to an external laser pulse. Further improvements, such as an increase in THz peak field or an extension of the IFEL interaction length, could push this THz streaking diagnostic to sub-femtosecond resolution.


From the comparison in Table~\ref{streakingcomparison}, we see that the greatest angular deflection for our current beam and undulator parameters is provided by the $\text{TE}_{02}$ mode. However, the simple fabrication of the offset mirror halves, as opposed to a THz waveplate, for coupling into the desired mode makes the $\text{TE}_{20}$ the most attractive method for implementing this THz-streaking concept at the PEGASUS laboratory.

\section*{Conclusion}

As the development of THz sources pushes towards higher power, the accelerator community eagerly looks to the potential of THz technology. To efficiently harness the power of a near single-cycle THz pulse for beam manipulation, the problem of extending the interaction region must be solved. Group velocity matching between a THz pulse and an electron bunch can allow for a sustained "zero-slippage" IFEL interaction. We intend to demonstrate this guided-THz IFEL technique at the PEGASUS laboratory to compress an electron bunch and reduce its time-of-arrival jitter with respect to the laser pulse generating the THz. Towards this end, we have achieved tunable control of the group velocity of a THz pulse inside a CPPWG. Initial predictions of the 1-D multifrequency simulation code that we are developing show bunch compression of up to an order of magnitude. The "zero-slippage" interaction that we utilize for compression can be extended to a transverse deflection mechanism, like the one discussed in ref. \cite{streakingAndonian}, without the need for an additional RF deflecting cavity. A THz streaking technique based on ponderomotive coupling in a guided-THz IFEL could provide unprecedented resolution for longitudinal bunch diagnostics. Looking forward, an FEL interaction driven by the guided-THz seed could prove to be a powerful tool for amplification of the THz source. 

\section*{Acknowledgment}
We are grateful for the support of the U.S. Department of Energy through grant DE-FG02-92ER40693 and the National Science Foundation through grant PHY-1415583 and acknowledge partial support from the U.S.-Israel Binational Science Foundation.

\section*{Appendix}

\footnotesize

\subsection*{THz source}

\begin{figure}[!htb]
   \centering
   \includegraphics*[scale=.3]{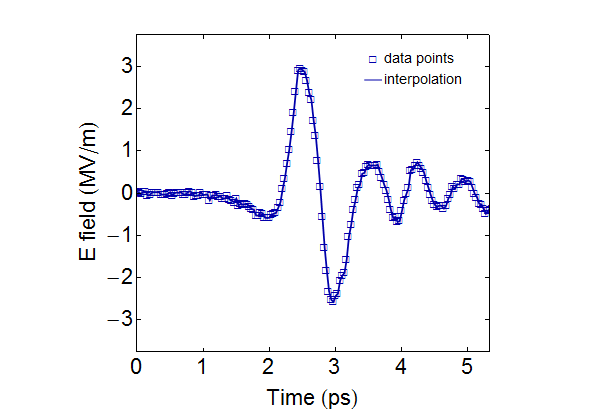}
   \caption{EOS measurement of the THz pulse at the entrance to the CPPWG.}
   \label{UDnoPPWG}
\end{figure}

We use optical rectification in stoichiometric lithium niobate (sLN) to generate picosecond scale THz pulses from a Ti:Sapph laser centered at $800\un{nm}$ with $30\un{nm}$ bandwidth. To enhance the conversion efficiency from IR to THz, the intensity front of the laser pulse is tilted using a diffraction grating and imaged onto the sLN with a focusing lens \cite{pulsefronttiltStepanov}\cite{pulsefronttiltHebling}. With $0.1\%$ conversion efficiency, we have $1.2 \,\mu \unx{J}$ pulses with a peak field of $3\un{MV}/\unx{m}$ after the diverging THz pulse is collimated and focused by a pair OAP's. A sample pulse profile is shown in Fig.~\ref{UDnoPPWG}. By increasing the laser spot size along with laser power, we can stay below the sLN damage threshold. Using $2.3\un{mJ}$ of IR power with doubled spot area, we produce a peak field of $4.6\un{MV}/\unx{m}$. We are currently developing a cooling chamber to further improve the conversion efficiency at the sLN crystal. With this modification and optimized focusing, we can expect the peak field to increase by at least a factor of two \cite{coolingBartal}.

\subsection*{EOS measurements}

Measurements of the temporal field profile are conducted with electro-optic sampling (EOS) in $0.5\un{mm}$ thick zinc telluride. Within this nonlinear optical crystal, a THz induced rotation of the fast and slow axes changes the relative intensity between the horizontal and vertical polarization components of an IR probe pulse. A balanced detection scheme, shown in Fig.~\ref{set_up_diagram}, utilizes a quarter waveplate and Wollaston prism to separate these components onto two photodiodes. The original THz field is then calculated in terms of the intensity difference \cite{numericalEOSBrunken}.

\bibliographystyle{unsrt}
\bibliography{/Users/scobywan/Dropbox/PBPL/Papers/theory_paper/v6/bibliofile}{}

\end{document}